\documentclass{iopart}
\usepackage{iopams,bm,epsfig}
\usepackage{graphicx}
\usepackage{exscale}
\usepackage{subfigure}
\usepackage{xspace}
\usepackage{citesort}

\newcommand{\<}{\langle}
\renewcommand{\>}{\rangle}
\newcommand{\be}{\begin{equation}}
\newcommand{\ee}{\end{equation}}
\newcommand{\bea}{\begin{eqnarray}}
\newcommand{\eea}{\end{eqnarray}}
\graphicspath{%
    {converted_graphics/}
    {C:/Users/Thila/Documents/PCTeX32/TexThila1/Malyshev/}
}
\begin{document}

\title{ Quantum information processing attributes of
 J-aggregates}

\author{ A. Thilagam} 
\address{Applied Centre for Structural and Synchrotron Studies, University of South  Australia, Adelaide,
South Australia  5095, Australia}
\date{\today}
\begin{abstract}
We examine the unique spectroscopic features which give rise to 
quantum information processing attributes of one-dimensional
J-aggregate systems, and  as revealed by entanglement measures such as 
the   von Neumann entropy,  Wootters
 concurrence and Wei-Goldbart geometric measure of entanglement.
The effect of dispersion and resonance terms in the
exciton-phonon interaction are  analyzed using Green function formalism
and present  J-aggregate systems as robust channels for 
large scale energy propagation for a select range of parameters.
We show that scaling  of  the third order optical response
$\chi^{(3)}$ with exciton delocalization size 
provides an experimentally   demonstrable  
measure of quantifying multipartite entanglement 
in J-aggregates.
 \end{abstract}
\maketitle
\section{\label{intro} Introduction}
Quantum J-aggregates in organic dye molecular systems 
reveal an unusually sharp and intense  red-shifted optical absorption band arising
from molecular  aggregation. This spectroscopic feature provides
 a spectacular example of nonlocal effects due to collective properties
of the Frenkel exciton \cite{jelly,sch,brigg1,brigg2}. 
The narrow isolated band of J-aggregates 
usually formed in liquid or solid organic matrix,
 generally appears adjacent to a  monomer based  broad absorption 
band incorporating vibrational features. The  absence of a similar 
vibrational structure in the polymer J-band is a striking
phenomenon, almost equivalent to a decoherence-free subsystem,
widely discussed in quantum information studies \cite{nel}. 

It is seen that subtle features of 
quantum J-aggregates   cannot be adequately
explained using conventional theory of exciton dynamics \cite{Dav,Craig,toy}.
Recent technical advances in spectroscopic techniques
such as two-dimensional (2D) coherent electronic 
spectroscopy with fast  femtosecond  time resolution \cite{expt1} 
has made it possible to observe intricate dephasing processes 
of delocalized exciton states in the single J-band aggregate 
system \cite{expt2}. Such detailed experimental probes of the
entangled properties highlight the potential in utilizing 
opto-electronics properties of J-aggregate
systems for quantum information processing. Optical-based 
applications have already been demonstrated in 
organic semiconductor microcavities \cite{micr}.
The potential application  to  innovative  ``frontier technologies" based 
on new understanding of the puzzling effects of quantum sized structures
therefore provide much impetus 
for further examination of J-aggregates, particularly
from a quantum information perspective. 

Narrow J-bands are generally studied  using a model  of 
 identical monomers interacting with vibrations \cite{sp,Klaf}
or an alternative model  based on  rigid monomers by just varying  electronic
site energies \cite{Knapp}.
The $SP$ parameter, given by
the  ratio of  inter-monomer interaction
energy to the vibrational width of the monomer  spectrum, 
 was  used by Simpson et al \cite{sp} to provide a measure of narrowness of the J-band.
In the strong coupling region with high $SP$ values, excitation transfer between
monomers occurs much faster than relaxation due to lattice vibrations.
In Knapp's   Frenkel exciton model \cite{Knapp}, 
specific details of the background medium was discarded 
with  attention focussed on the  the J-band. The time 
of propagation of the  Frenkel exciton was considered 
faster than time variations in electronic site  energies.
The term ``exchange narrowing" \cite{Knapp}
was used to describe the decrease 
in linewidth  by $1/\sqrt{N}$  in
the absence of  intersite correlation, $N$ being the number of monomers. 
The  "exchange narrowing" process can be associated with
 $N$ monomers  participating  
 in a  collective mode as a  multipartite system,
 each monomer thus  experiencing  
an effective disorder of magnitude $\frac{D}{\sqrt{N}}$ instead of
$D$, the full disorder experienced by an isolated monomer.

It is well known that quantum non-locality is a collective property 
of  quantum  system of two or more subsystem (e.g monomers)  in which the 
dynamics of constituent subsystems depend on   
spatially separated  adjacent subsystems \cite{nel}.
Pauli exclusion principle excludes the possibility of two
excitons occupying the same molecular site, thus
processes such as exciton-exciton scattering and excitation/de-excitation 
at a site invariably
leads to entanglement in linear excitonic chains. 
The ``spooky action at a distance" effect \cite{spookE} ensures 
minimal participation of vibrational modes in quantum J-aggregates.
The time taken for individual monomers to
interact and form the collective J-band is far less than interaction
time with lattice vibration as revealed experimentally \cite{expt1,expt2}. 
Recently the puzzling effects of 
entanglement was experimentally demonstrated \cite{spookJ} by
joint weak measurement procedures using  an entangled
photon pair.  

In this work, we examine the dynamics  of J-aggregates using 
a quantum information approach in which 
the  origin of its sharp J-band is attributed 
to   entanglement and non-local effects 
associated with a system of excitonic monomers. We consider
that decoherence effects \cite{tiop1,tiop2,tiop3,briggy} associated with a noisy 
environment invariably  disrupts the delicate 
delocalized excitation  which contributes to the narrowness
of  J-aggregate spectrum. 
In  the extreme limit of an infinite chain with infinitely strong coupling, we
assume that nonlocal effects  is  maximal  with characteristics
of an ideal J-band  forming  an infinitely narrow spike.

J-aggregates appear remarkably to be the first systems in which 
 entanglement effects have been  revealed so visibly 
during the 1930's,  understandably  
Scheibe \cite{sch} attributed the J-band as arising due to  a collective
state of electronic excitation.  Despite the consistent observation of a collective
excitonic mode  by Scheibe \cite{sch},  Simpson et al \cite{sp} and Knapp \cite{Knapp}, attempts to
bridge the gap between the well-known Frenkel exciton theory and the more
 recent field of quantum information theory 
has only been achieved recently in our work \cite{ThilaPRA},
where we have shown clear and explicit links between the quantum search process
and actual physical processes taking place in the crystal.
To simplify the numerical analysis in this work, we utilize the Markovian approximation scheme 
in which  the  time scales of lattice vibrations are considered to be short compared to
other dynamical times present in the system. Hence we 
neglect the   flow of information from phonons back to the excitonic system
and assume that the reservoir system is disengaged from 
coherence effects between two or more molecular sites. In other words, we consider that
the operating times  associated with non local effects remain {\it unaffected} by ``memory effects"
present in  a non-markovian reservoir system. We thus distinguish between propagation times of the
excitation and times associated with  entanglement effects  of a delocalized exciton which
determines the duration of coherence between different molecular sites.

In view of the importance of including  entanglement effects in 
J-aggregates, we  emphasise  the general problem 
of quantifying the level of entanglement in any arbitrary multipartite system.
 Currently there is no satisfactory
classification of genuine entanglement for a multipartite system of even pure states
 \cite{scott,hor,ami,hein,plen,eise}. The manner of partitioning 
individual subsystems affects  multipartite entanglement and introduces an
element of subjectivity when using  current  classification of entanglement such
as k-separability. For instance  for a selected partition, one can obtain varying degrees
of multipartite entanglement, changing  from a
fully separable to  fully inseparable states with  each
member of the group  sharing  genuine multipartite entanglement. The possibility that
such a change may be non-monotonic was recently shown in an earlier work
where three members of a group which shared non-zero tangle entanglement displayed 
zero  bipartite entanglement between  two members \cite{coff}. We therefore expect
investigations related to  maximal entanglement of a fragile open systems
such as  excitonic chains with decoherence effects 
 to remain unresolved  for some time. In this regard, we point 
to the possibility that the  scaling  of  the third order optical response
$\chi^{(3)}$ with exciton delocalization size  may provide
an experimentally   demonstrable means of categorizing  
genuine  multipartite entanglement, at least in the case of J-aggregates.

This paper is organised as follows. In Sec. \ref{Theory}, we summarize 
salient features of the Frenkel exciton in molecular systems, including
a brief description of the site basis representation appropriate 
to excitonic systems.  In Sec. \ref{quantent}, we provide a brief
description of the von Neumann entropy \cite{ami,kit}, and 
in Sec. \ref{time} we provide mathematical details of
derivations based on  Green's functions formalism. We
also present results which show the usefulness of 
the entropy in  capturing the transition from coherent to incoherent energy transfer
in a finite one-dimensional J-aggregate system. In Sec. \ref{cycle}, we use the 
 well-known  Wootters  concurrence \cite{Woo} to  interpret  
superradiance properties of one-dimensional 
excitonic systems. Finally, we analyze  the intricate link between 
 the third order optical response
$\chi^{(3)}$ and  the class of 
entanglement quantity known as the Wei-Goldbart 
geometric measure of entanglement \cite{Wei2} in  Sec. \ref{opt}.

\section{J-aggregates and Frenkel Excitons }
\label{Theory} 

J-aggregates, also called ``communal" states \cite{sch},
are low dimensional molecular systems which display enhanced
coherent interactions with light. For instance, a well known dye molecule
pseudoisocyanine (PIC) forms a quasi-one dimensional arrangement
of molecules under suitable conditions,  displaying the 
characteristic feature of a sharp absorption spectrum red shifted from the broader 
monomer transition band. It is well known that many spectroscopic features of J-aggregates
are interpreted via a model of one-dimensional Frenkel exciton system \cite{Dav},
with delocalised excitons considered the main reason behind the
collective nature of such systems. There have been suggestions that the 
total number of molecules which form such collective states varies from
ten to several hundreds \cite{Knapp}. 

The optical properties of J-aggregates are invariably linked
to geometrical attributes of the molecular arrangement, however in this
work we restrict our study to 
Frenkel excitons in a quasi one-dimensional system
of finite $N$ molecular sites. We assume that the
one-dimensional Frenkel excitonic model is applicable to 
 one-dimensional aggregate systems as well.
Excitons in molecular crystals are generally
treated using the   tight-binding model \cite{Dav}
due to the weak  overlap of intermolecular
wavefunctions and hence localization of excitation 
at individual molecular sites.
 The exact form of these functions are not needed to derive salient
properties of Frenkel excitons, however every molecular site 
has an equal probability of being excited by an incident
photon due translation symmetry, a salient feature which  
has important implications for  quantum parallelism.
Various mechanisms result in the loss of excitonic coherence,
in this work we consider the effects due to coupling with
lattice vibrations as well as the influence of  finite temperature.

The absorption of a photon  triggers an initialization
 process with creation of a delocalized
 exciton,  representing an equal weighted superposed form
of occupation probabilities  at all lattice sites.  While
ideal crystal in vacuum state is equivalent to the
null state $|0\>^{\otimes N}$ for $N$ sites, 
photon absorption process initiates
a Hadamard-like transformation \cite{nel} at each site. In the absence of
lattice vibrations, the exciton  eigenvector can thus be written as
\be
\label{ex}
|{ K} \> =  N^{-1/2} {\sum_{ l}}
e^{i { K . l}} B_{ l}^\dagger |{ 0} \> 
\ee
where the reciprocal lattice vector  $K= \frac{2 \pi k}{N}$ within
the Brillouin zone $k \in [-\frac{N}{2},\frac{N}{2}]$,
$|{ 0} \> $
is the vacuum state with all molecules in ground state
 and 
$B_{ l}^\dagger$ is the creation operator of exciton
at position coordinate  ${ l}$. $N$  
is the number  of  unit cells in the crystal which increases
with the number of molecules for each unit cell. 
In Eq.(\ref{ex}), we have dropped notations associated with the exciton
spin for convenience reasons,  and include  notations in scalar form
 as we restrict  the model considered here  to a finite 
one-dimensional linear chain. 
It should be noted that we have considered that all molecular sites have
an equal chance of absorbing a photon to raise it to an excited state $|{ 0} \>$ 
which can be represented by the Hadamard-like transformation: $|{ 0} \>  \rightarrow |0\>+|1\>$.  
It is important to note the vital difference between a photon absorption process
and a true  Hadamard gate conversion which can reset a state to its original form.
For instance, other possible transformations inherent in a true Hadamard gate such
as $|0\>+|1\>  \rightarrow |0\>$ and $|1\>  \rightarrow |0\>-|1\> \rightarrow |1\>$
are not  present upon successive 
absorption of two or more photons  at a molecular site.

The creator operator of a Frenkel exciton with wavevector ${ K}$
can be easily obtained using the Fourier transform of Eq.(\ref{ex})
\be
\label{exF}
B_{ K}^\dagger =  N^{-1/2} {\sum_{ l}}
e^{i { K.l}} B_{ l}^\dagger
\ee
The exciton creation operator $B_{ K}^\dagger$ localized in $K$-space
is delocalized in real space \cite{Dav}. The motion of
the exciton in molecular systems is governed by an Hamiltonian
derived using a tight-binding model \cite{Dav,Craig,toy}
\be
\label{exH}
\hat{H}_{ex}= \sum_{ l}\left( {\Delta E} +
{\sum_{ m \neq l}} D_{ l,m} \right )B_{ l}^\dagger B_{ l}+
\sum_{ m \neq l} M_{ l,m} B_{ l}^\dagger B_{ m}
\ee
where $\Delta E$, the on-site (intra-site) excitation energy at equilibrium 
is the same at all sites due to translational symmetry. 
$D_{ l,m}$ is the dispersive interaction matrix element
which determines  the   energy difference between a pair of 
excited electron and hole at a molecular site and ground state electrons
at neighboring sites \cite{Dav,Craig}.  $M_{l,m}$  the 
electron transfer matrix element between molecular sites at ${ l}$
and ${ m}$. Using Eq.(\ref{exF}), we obtain a Hamiltonian 
 diagonal in ${ K}$ space
\bea
\label{exHk}
\hat{H}_{ex} &=& \sum_{ k} E_0({ k}) B_{ k}^\dagger B_{ k} \\
\nonumber
E_0({ k})&=&\Delta E+\sum_{ m \neq 0} D_{ 0,m}+ M_{ 0,m} \exp(i{ k.m})
\eea
where $E({ k})$ is energy of the exciton of wavevector ${ k}$ and subscript
$0$ denotes the absence of lattice site fluctuations.
For a simple one-dimensional lattice system, we approximate
 $M_{l,m} \sim V_{i,j}$ where $V_{i,j}$,  the neighbor transfer energy is given by 
\be
\label{transJ}
V_{i,j} \propto \frac{\mu_i \mu_j}{d^3}
\ee
$\mu_i$ the molecular transition moment at site $i$
and $d$ is the separation between the two sites involved
during the energy transfer.

\subsection{\label{site} Site basis representation}

There are several ways to represent excitonic states as qubit states.
For instance, Eq.(\ref{ex}) may be considered as the result of a single
step of a naturally occurring process involving photon absorption,
requiring the equivalent of $\log_2(N)+1$ qubits to store the superposed state.
On the other hand, each molecular site may be occupied by an exciton 
and thus considered a qubit depending on the presence (or absence) of an exciton at a
particular site and hence a two-level system.  Eq.(\ref{exH})
can be formulated in terms of the terms of the exciton density matrix as:
\be
\label{exdensity}
\rho(t) = \sum_{n=1}^N \rho_{n n}(t) |n\>\<n| 
+ \sum_{n<m}^N  \rho_{n m}(t) |n\>\<m|  + \rho_{n m}^* (t)|m\>\<n| 
\ee
where $|n\>$ represents a site basis state    where the exciton is present at only 
the $n$th molecular site with all other sites unoccupied. The    basis state
 $|n \>$ denotes  $|0\>^{\otimes (n-1)}\otimes|1\>\otimes|0\>^{\otimes (N-n)}$, where  $N\times N$
is the dimension of  the  global density matrix. $|0\> (|1\>)$ denotes 
the absence (presence) of an excitation at a given molecular site.
The  $N$ molecular sites occupied by the exciton  correspond to the $2^N$ discrete states in the Hilbert space of 
dimension $N$. Transitions between the states
in Hilbert states hence correlate with  the transition
amplitudes in the original Hamiltonian in Eq.(\ref{exH}).

 A  qubit state $\Psi$ associated with $N > 2$ subsystems can be written as
generalized Schmidt Decomposition belonging to the Hilbert space of $\hat{H}_{ex}$ in
 Eq.(\ref{exH})
\be
|\Psi\> = \sum_{n=1}^N c_{n} \left( |0\>^{\otimes (n-1)}\otimes|1\>\otimes|0\>^{\otimes (N-n)} \right)
\ee
 with Schmidt coefficients  $c_n$. 
The Schmidt Decomposition above is well known  for the
 separability characterization of pure states.
For the simpler case of just two molecular sites $N=2$, the bipartite
 is separable only if there is  one non-zero Schmidt coefficient. 
The state can be considered  maximally entangled if 
 all the Schmidt coefficients are non-zero and equal.
The well known Dicke state 
$|W \> =\frac{1}{\sqrt{N}} \left(|100...0\>+|01...0\>+...+|0...01\> \right)$
is  symmetric under permutations of each basis state $|n \>$. 
The absorption of a single photon will give rise to a superposition of all
 possible states as shown in 
 Eq.(\ref{ex}). These collection of states  will also include the special subset of 
Dicke states which constitute  an equal weighted superposed form
of  basis states.
 
\section{\label{quantent} Quantifying entanglement via the von Neumann entropy}

We use the  local density
matrix $\rho_{n} = u_n |1 \>_n {}_n\<n|+(1-u_n) |0\> \<0|$ 
to determine  entanglement dynamics of the $n$th qubit 
associated with a basis which spans the two dimensional space 
($|0\>$  and $|1\>$). 
The occupation probability at the $n$th molecular 
site is given by $u_n =  \<{ K}|B_{n}^\dagger B_n |K\>$
where $B_n$ is the exciton creation operator 
localized at site $n$ and $|K\>$
is  the exciton  eigenvector in  $K$-space (see Eq.(\ref{ex})).
The reduced density matrix
has a simple form  
\be
\label{dm}
\rho_{n} = \left(\begin{array}{clcr}
1-u_n  &  0\\
0  &   u_n \\
\end{array}
 \right).
\ee

A suitable entanglement measure for the Frenkel exciton described
by the one-dimensional   tight-binding model in Eq.(\ref{exH})
and corresponding to the density matrix in Eq.(\ref{dm})
is the widely used von Neumann entropy measure \cite{ami,kit}
\be
\label{entrop}
S_n = -{\rm Tr}_n \rho_n \ln \rho_n
\ee
where $S_n$ which is zero for pure states, 
is obtained by tracing over the degrees of freedom
of the remaining $N-1$ molecular sites. The average 
von Neumann entropy measure is then given by 
\be
\label{vN}
{S} = \frac{1}{N}
\sum_{n=1}^{N}  -{\rm Tr}_n \rho_n \ln \rho_n
\ee
For the ideal delocalized state
in Eq.(\ref{ex}), we obtain ${S} = \frac{1}{N} \ln N$ at very large 
$N$. The larger the entropy,  greater is the
entanglement between excitation at a given site with all 
 other remaining sites.
On the other hand, $S=0$ for the localized state and hence
is a useful measure in characterizing  quantum phase
transitions at the localization-delocalization borders. 
For the case of an extended state in a finite system of size $N$,
we obtain ${S} = \frac{1}{N} \ln N - (1-\frac{1}{N}) \ln (1-\frac{1}{N})$.
Using a time dependent density matrix (as will be considered in the next Section), we will
show that the time evolution of the corresponding 
von Neumann entropy measure can reveal variations in 
 entanglement properties as the exciton propagates in a noisy
environment.

 \section{\label{time} Green's function formalism and probability propagator}
We begin with the  following crystal Hamiltonian  \cite{Dav}
\bea
\label{exHim}
\hat{H}_{T}&=& \hat{H}_{ex} + \hat{H}_{p}
+\hat{H}_{ep1} + \hat{H}_{ep2}\\
\label{phon}
\hat{H}_{p}&=&\sum_{ q} \hbar \omega({ q})b_{ q}^\dagger b_{ q}
\\
\label{exphon1}
\hat{H}_{ep1}&=& N^{-1/2}\sum_{ k,q} \left[F({ k, q}) + \chi({ q}) \right]B_{ k+q}^\dagger
B_{ k} (b_{- q}^\dagger +b_{ q}) \\
\label{exphon2}
\hat{H}_{ep2}&=&
N^{-1/2}\sum_{ k,q} \chi({ q}) B_{ k}^\dagger B_{ k} (b_{- q}^\dagger +b_{ q}) 
 \eea
where $\hat{H}_{ex}$ is given in Eq.(\ref{exHk}), $\hat{H}_{p}$ denotes the
phonon energies and $b_{ q}^\dagger (b_{ q})$ is the creation (annihilation) phonon
operator with  frequency $\omega({ q})$ and  wavevector ${ q}$. Explicit forms for the
coupling functions $F({ k, q})$ and $\chi({ q})$ are given in Ref. \cite{Dav}. It is important
to note the difference in the two coupling functions, while  $\chi({ q})$ acts
to localize an exciton at its original occupation site,  $F({k, q})$  operates during
dispersion of an exciton to neighboring sites. 
Thus $F({ k, q})$ is influenced by the same factors which
determine the exciton bandwidth. In situations where $\chi({ q}) >> F({ k, q})$,
self-trapped  excitonic effects are expected to dominate while in the case
of comparable coupling strengths $\chi({ q}) \sim F({ k, q})$, 
coexistence of free and trapped excitons and localization effects are expected to prevail.
Due to the finite time associated with lattice vibrations,
nonlocal effects which govern  entanglement properties of the delocalized exciton
are likely to experience decoherence effects due to the first coupling term 
in Eq.(\ref{exphon1}). At very high neighbor transfer energies  $J$,
the  second dispersive term in Eq.(\ref{exphon2}) acts as a small
perturbation on   exciton dynamics. We now investigate the effect of these
different  coupling terms using Green's functions formalism introduced by Davydov \cite{Dav}.

The Green's function for an exciton at time
$t$ is given by \cite{Dav,suna}
\bea
\label{g1}
G({ k},t)&=&-i \<0:n_{ q}|T\{B_{ k}(t) U(t,0)B_{ k}^\dagger(0)\}|0:n_{ q}\>
\\ \label{ge}
B_{ k}(t)&=&B_{ k}(0) \exp(i E({ k})\; t)
\eea
where $|0:n_{ q}\>$ denotes a state with 
$n_{ q}$ as the  population  of phonons with wavevector ${ q}$
and zero exciton population. $T$ is the time ordering operator and
$U(t,0)$ is the Dyson function
\be
\label{Dy}
U(t,0)=T \exp\left(-i \int_0^t \exp(i \hat{H}_{T} t') \left[\hat{H}_{ep1}+\hat{H}_{ep2} \right]
\exp(-i \hat{H}_{T} t') dt'\right)
\ee
where $\hat{H}_{T}$, $\hat{H}_{ep1}$ and $\hat{H}_{ep2}$ are given in  Eqs. (\ref{phon}),
(\ref{exphon1}) and (\ref{exphon2}) respectively.
A direct relation between the Green's function $G(k,t)$ in 
Eq.(\ref{g1}) and exciton energy  $E({ k})$ has been
derived by Craig et al (see
Eqs. 38 and 39 in Ref. \cite{craigdiss})  at times larger
than the inverse of phonon frequencies,
\bea
\label{g2}
G(E({ k}) ,t)&=& -i \exp\left[-i \; \left(E({ k})+ \Delta({ k}) -
N^{-1} \sum_{ q} \frac{|\chi({ q})|^2}{\omega({ q})^2}\right)\; t \right]
\\ && \times
\exp[-\gamma({ k})] \exp(-\Lambda)
\eea 
where 
 \bea
\label{g3}
\Delta({ k})&=&N^{-1}\sum_{ q} |F({ k,q})|^2 \left[\frac{\bar{n}_q}{\Omega^-_{ k,q}}
-\frac{\bar{n}_q+1}{\Omega^+_{ k,q}} \right ]\\
\label{g4}
\gamma({ k})&=& N^{-1}\sum_{ q} |F({ k,q})|^2 \left[\bar{n}_q \; \delta(\Omega^-_{ k,q})
+(\bar{n}_q+1)\; \delta(\Omega^+_{ k,q}) \right ]\\
\Omega^{\pm}_{ k,q}&=&\hbar \omega({ q})\pm(E({ k+q})-E({ k}))\\
\Lambda&=&N^{-1} \sum_{ q} \frac{|\chi({ q})|^2}{\omega({ q})^2} (2 \bar{n}_q+1)
 \eea
where  $\Delta({ k})$ yields the principal value of the term associated with it,
and the mean phonon occupation number at temperature $T$ is given by 
$\bar{n}_q =\{ \exp[\hbar \omega({ q})/k_B T]-1\}^{-1}$. We note that
$\Lambda$ is independent of time unlike the other terms, however this 
only applies in the Markovian approximation. A rigorous
treatment in the non-Markovian range involves intense numerical computations and 
therefore will not be considered here. Nevertheless  the 
Green's function $G(k,t)$ in  Eq.(\ref{g2}) will suffice to reveal
vital difference in the roles played by the  two coupling
functions,  $\chi({ q})$ and $F({ k, q})$ in influencing the times of coherence
 between lattice sites.

The dynamics of a free particle
in a  one-dimensional system is determined entirely
by the Green function $G_{n,m}(t)=\<n|G(t)|m\>$  \cite{econ}.
which  yields the amplitude for an excitation
to move from site $m$ at $t=0$  to site $n$ 
at time $t$. In a one-dimensional lattice 
with nearest-neighbor interaction energy $2 V \cos(K)$ 
and zero  lattice vibrations,  $G_{n,0}(t)$ 
is given by
\be
\label{econ1}
G_{n,0}(t)= N^{-1}\sum_{ k} e^{i k n} \exp \left[ -i \frac{2 V t}{\hbar} \cos(K) \right ]
\ee
Accordingly, the probability $P_{n0}(t)$ that an excitation which 
originates at site $0$ at $t=0$  and appears at site $n$ 
at time $t$ can be obtained  using 
$P_{n0}(t) = |G_{n,0}(t)|^2$. 
To calculate the probability $P_{n0}(t)$ for the exciton
in the presence of dispersion and resonance terms 
$\chi({ q})$ and $F({ k, q})$ quantifying
the exciton-phonon coupling, we use the Green's function
$G(E({ k}),t)$ in Eq.(\ref{g2}). For very large $N \rightarrow \infty$,
we obtain
\be
\label{pron}
P_{n0}(t) \approx N^{-1}{\rm J}_n^2(\frac{2 V}{\hbar} t)\exp(-\Lambda)\sum_{k} e^{- \gamma(k) t}
\ee
 where ${\rm J}_n(x)$ is Bessel function of the first kind and $\gamma(k)$ is given
in  Eq.(\ref{g4}). In the absence of lattice vibrations, $P_{n0}(t)$
reduces to the well-known form obtained by Merrifield \cite{merry}
for an  excitation propagating on infinite aggregate systems.
 We have assumed that the transfer interaction term $V$ is independent
of the environmental variables associated with lattice vibrations, also known as
Condon approximation in Eq.(\ref{pron}).
Recent experimental results \cite{expt1} have shown 
that  electron-phonon interactions between  off-diagonal elements
modulate the couplings between the dipoles, resulting in decoherence of 
the transfer interaction term $V$. Here we assume that 
the transfer interaction term $V$ remains intact, and 
Eq.(\ref{pron}) is a reasonably accurate description of the exciton propagation 
 in the presence of lattice vibrations. 

The  subtle differences in influence of the coupling
functions,  $\chi({ q})$ and $F({ k, q})$
on the probability transfer function $P_{n0}(t)$  is evident
from Eq.(\ref{pron}). The dispersive coupling factor 
 $\chi({ q})$   tends to localize the excitation at the original
site of excitation  rather than introduce
decoherence during propagation. Oh the other hand,
 the resonance coupling term  $F({ k, q})$
 acts to retard the tendency for the  exciton to move further away from 
the initial site of propagation by contributing to decoherence effect {\it during}
propagation. Eq.(\ref{pron}) can be rewritten in terms
of dimensionless arguments
\be
\label{pron2}
P_{n0}(t) \approx N^{-1}{\rm J}_n^2(c t)\exp(-a^2)\sum_{k} e^{- b^2 t}
\ee
where $c = \frac{2 V}{\hbar  \omega({ q})}$, 
$a = \frac{\chi({ q})}{\hbar  \omega({ q})}$ and 
$b = \frac{F({ k, q})}{\hbar  \omega({ q})}$. Time $t$ is 
dimensionless as it is scaled by the factor $ \omega({ q})^{-1}$. It is to be noted that
 $t> 1$ describes the Markovian limit reasonably well. 
At the Markovian limit,  the flow of information from phonons back to the excitonic system 
can be neglected as the exciton hopping times are large compared to the 
phonon reservoir correlation times. This means that two propagating 
exciton states cannot be distinguished as they remain unaffected by  correlations within
the phonon reservoir at times larger than phonon correlation times.

In order to obtain an explicit result from based on Eq.(\ref{pron}), we make a few approximations,
 firstly we neglect the dispersion in  phonon frequencies so that a single frequency, 
 $ \omega({ q}) \sim 50$cm$^{-1}$,  representative of the lowest vibrational
branch in common molecular crystals \cite{craigdiss,paw} is considered. 
The band averages of the two  coupling
functions,  $\chi({ q})$ and $F({ k, q})$ are then varied via parameters $a$
and $b$ (given below Eq.(\ref{pron2})),
together with the electronic transfer energy, quantified by $c$.
As mentioned earlier, we consider a system of a finite number of molecular sites $N$,
and evaluate the  von Neumann entropy measure of the system using 
\be
\label{vs}
{S} =  \sum_{n=1}^{N}  -{\rm Tr}_n \rho_n \ln \rho_n
\ee
The influence of the coupling functions obtained
using Eq.(\ref{pron2})  
on the  von Neumann entropy measure given in  Eq.(\ref{vs}) can thus be easily
evaluated. 

\begin{figure}[htp]
\begin{center}
    \subfigure{\label{fig1a}\includegraphics[width=6.0cm]{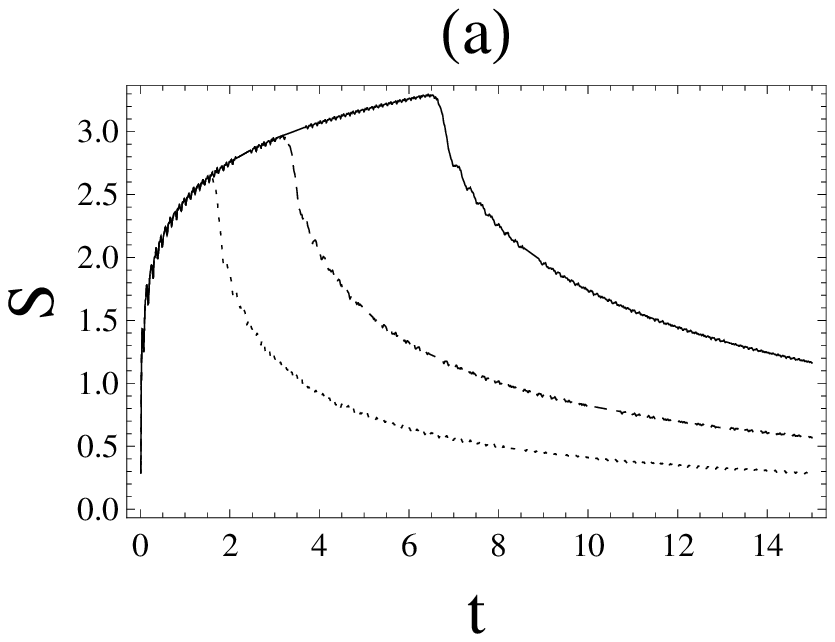}}\vspace{-3.0mm} \hspace{-1.1mm}
    \subfigure{\label{fig1b}\includegraphics[width=6.0cm]{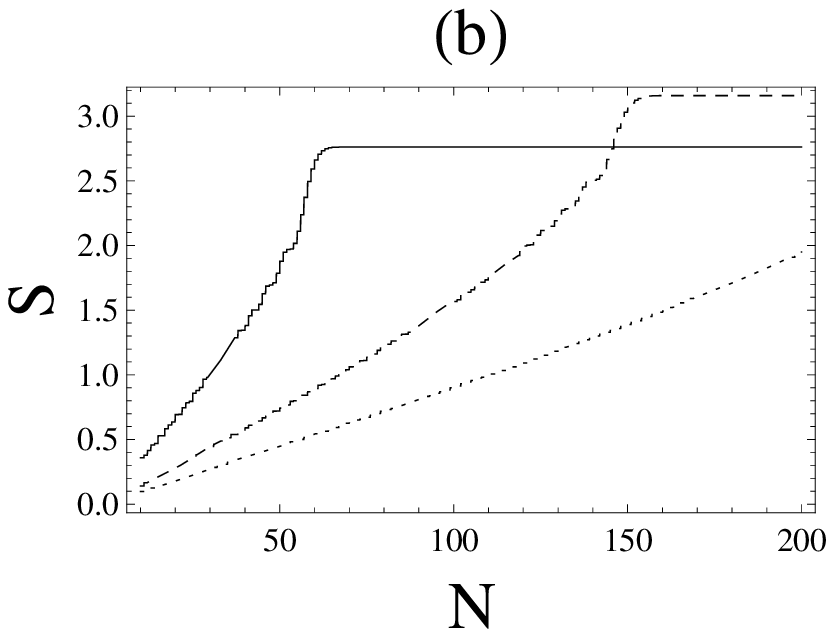}}\vspace{-4.0mm} \hspace{-1.1mm}
    \subfigure{\label{fig1c}\includegraphics[width=6.0cm]{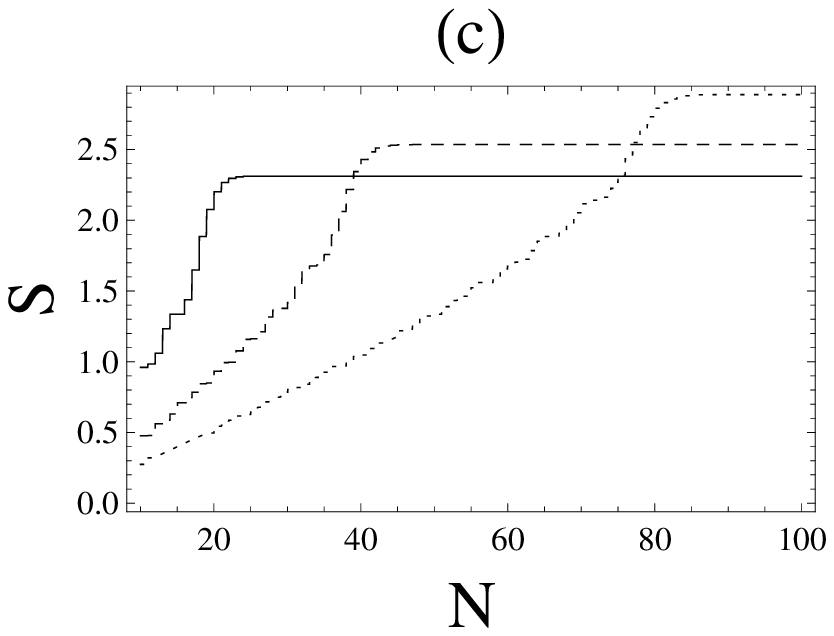}}\vspace{-4.0mm} \hspace{-1.1mm}
    \subfigure{\label{fig1d}\includegraphics[width=6.0cm]{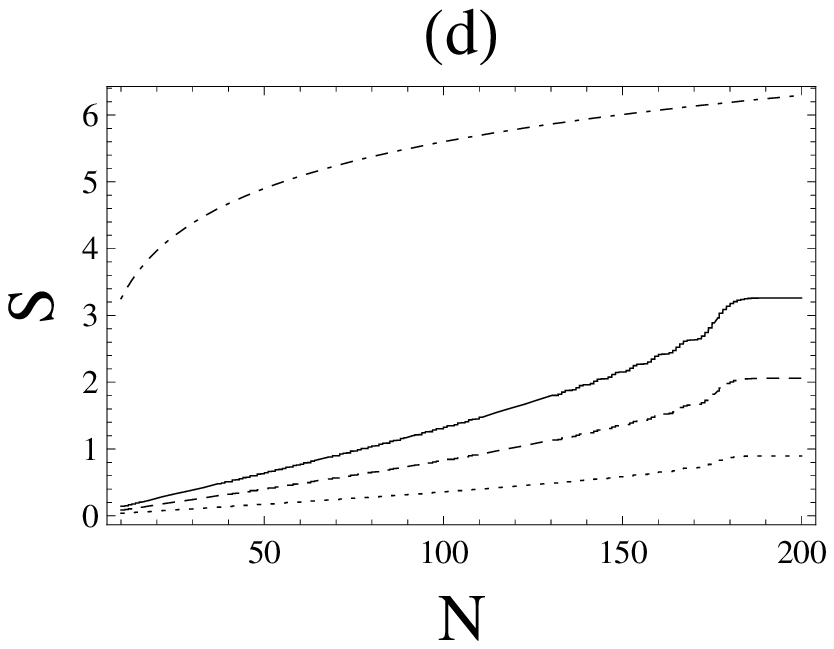}}\vspace{-4.0mm} \hspace{-1.1mm}
  \end{center}
\caption{\label{w1}
a) The von Neumann entropy $S$ versus time $t$ at $N=200$ (solid line),
$N=100$ (dashed)  and $N=50$ (dotted) at $a=b=0$ and $c=30$.
\\
b) The von Neumann entropy $S$ versus time $N$ at $t=2$ (solid line),
$t=5$ (dashed)  and $t=9$ (dotted) at $a=b=0$ and $c=30$.
\\
c) The von Neumann entropy $S$ versus time $N$ at $c=10$ (solid line),
$c=20$ (dashed)  and $c=40$ (dotted) at $a=b=0$ and $t=$2.
 \\
d) The von Neumann entropy $S$ versus time $N$ at $b=0$ (solid line),
$b=0.3$ (dashed)  and $b=0.5$ (dotted) at $a=0$, $c=30$ and $t=6$. 
Dotted-dashed line corresponds to 
a fully extended state for which ${S} = \frac{1}{N} \ln N - (1-\frac{1}{N}) \ln (1-\frac{1}{N})$
}
\end{figure}

In Fig.~\ref{w1}a, we show  results of $S$ versus time $t$
obtained using  Eq.(\ref{vs}) for various values of $N$, and 
for an initial state, $P_{n0}(0)= \delta_{n,0}$.
At time $t=$0, average entropy $S=$0 
due to presence of excitation at just one site, however the entropy rises rapidly  
reflecting the spread of excitation to more sites  with
constant velocity (determined by $c$). 
Infact $S$ appears to display an approximate linear dependence on $N$, as it should
be as a system with large $N$ results in the occupation of more lattice sites and greater 
chance of existing as an extended system.
However as we are considering a finite system, the propagating exciton leaves  the system 
under study  after a critical time. Hence the entropy $S$ can be seen to drop  in all three cases of $N$ considered
in Fig.~\ref{w1}a. This drop is most pronounced at small $N$ where there faster exit of the exciton.
It is to be noted that both parameters 
$a$ and $b$ are set to zero in Fig.~\ref{w1}a so 
that phonon-coupling processes are absent, and this
confirms that the  noticeable drop in $S$ is indeed due to the finite size effects of our chosen model.

It is important to note that at small time $t$, there is a limit to the
number of sites visited by the exciton and thus we expect an upper limit to $S$ which is 
noticeable at $t \sim 2$. To illustrate this point, 
we show  results of $S$ versus 
number of sites $N$ at times $t=$2 (solid line), $t=$5 (dashed)   and $t=$9 (dotted) in 
Fig.~\ref{w1}b. We note that at small $t$, entropy $S$ becomes flat after propagation 
to a finite number of sites.  Thus the entropy $S$ increases with $N$ as more sites are 
visited before becoming independent of $N > N^*$. The critical  $N^*$ is  
is  dependent on  $c$ and hence the
transfer interaction term $V$. At larger $t$, more sites have a non-zero probability of 
occupation and therefore $N^*$ increases with $t$. 
 The dependence of $N^*$ on speed of excitation transfer (quantified by $c$)
 is illustrated in Fig.~\ref{w1}c where we note increase in $S$ and 
$N^*$ with increasing $c$. The effect of the phonon coupling function
$b = \frac{F({ k, q})}{\hbar  \omega({ q})}$ on $S$ is illustrated in 
Fig.~\ref{w1}d, which shows the gradual decrease of $S$ due to increase 
coupling with phonons. Figs.~\ref{w1}a-d thus illustrate clearly  
the interplay of three competing effects: 1)  the increase of $S$ due to 
spreading; 2) subsequent decrease of $S$ due to finite size effects 
(propagation out of the system) and, 3) decrease of $S$ due to  coupling with phonons.

\begin{figure}[htp]
  \begin{center}
    \subfigure{\label{fig1}\includegraphics[width=7.0cm]{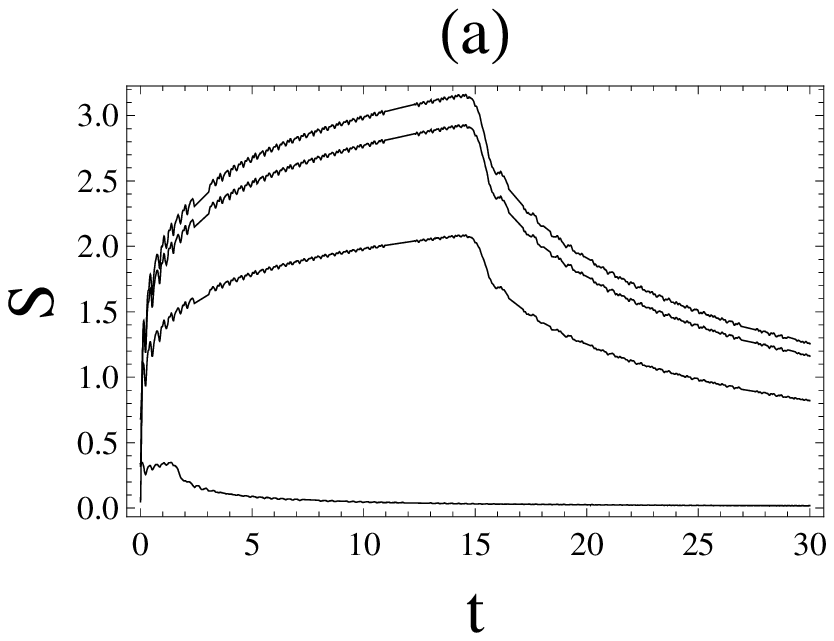}}\vspace{-3.0mm} \hspace{-1.1mm}
    \subfigure{\label{fig2}\includegraphics[width=7.0cm]{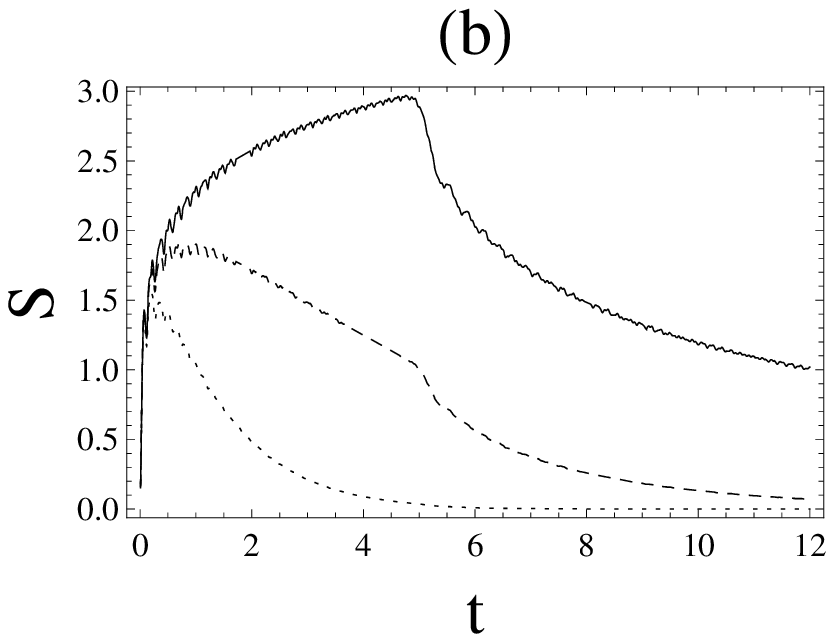}}\vspace{-4.0mm} \hspace{-1.1mm}
    \subfigure{\label{fig3}\includegraphics[width=7.0cm]{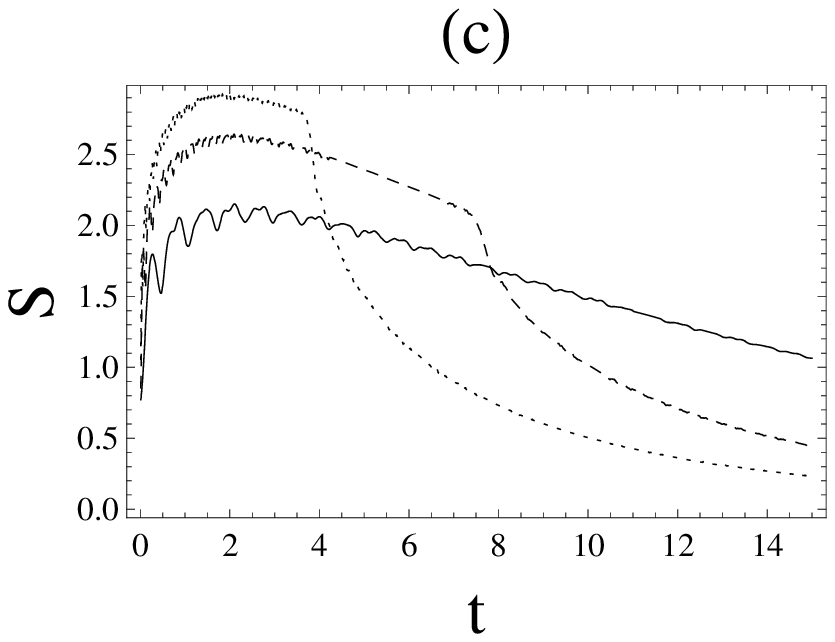}}\vspace{-4.0mm} \hspace{-1.1mm}
  \end{center}
  \caption{a) The von Neumann entropy $S$ versus time $t$ at $N=150$, $b=0$  and $c=$10.
The four curves correspond to (from top to bottom) $a=$ 0, 0.3, 0.7 and 1.5. \\
b) The von Neumann entropy $S$ versus time $t$ at $N=100$, $a=0$  and $c=$20. 
The three curves correspond to (from top to bottom) $b=$ 0, 0.5 and 1. \\
c) The von Neumann entropy $S$ versus time $t$ at $N=200$, $a=0.5$  and $b=0.3$.
The three curves correspond to $c=$ 40 (dotted), 20 (dashed) and 5 (solid line).
}
 \label{wfig}
\end{figure}

In Fig.~\ref{w1}d, we have included a dotted-dashed line which corresponds to 
a fully extended state at which ${S} = \frac{1}{N} \ln N - (1-\frac{1}{N}) \ln (1-\frac{1}{N})$.
The entropy $S$ of the extended state is higher than other $S$
values calculated for the case of the one-dimensional system considered in Figs.~\ref{w1}a to d.
The results indicate that   at large $N$, the entanglement persist for longer times
and  $S$ is almost comparable to that of a fully extended system. 
This suggests  the large scale energy transport abilities of one-dimensional excitonic
systems. We next examine in detail the  mechanisms associated with varying coupling parameters 
$a$ and $b$ next ( Fig.~\ref{wfig}a,b,c.)

Results obtained by varying $a,b,c$ 
is illustrated in Fig.~\ref{wfig}a,b,c and imply
 agreement with localization tendencies associated with coupling 
parameters, $a$ and $b$. The fall in $S$ is noticeably steep for 
high values of $b \sim 1$ at  intermediate  $N \sim 100$. This shows the higher 
localization properties associated with  the  coupling
function  $F({ k, q})$ compared to $\chi({ q})$ (see Fig.~\ref{wfig}b).
Fig.~\ref{wfig}c illustrates  that high  electronic transfer energies or  $c$
lead to fast rise in $S$ in the time domain ($0 < t < 6$), which however is not
sustained as excitation leaves the system as quickly as it spreads. For lower
$c$, the excitation remains extended for comparatively longer times with noticeable
oscillations associated with ``death and birth" type excitation and de-excitation process \cite{thilaIOP}
  at a monomer site. The results obtained in Figs.~\ref{w1} and ~\ref{wfig}a,b,c
highlight the prominence of the  J-band in large systems and show the usefulness of 
the entropy in  capturing the transition from coherent to incoherent energy transfer.

So far we have only considered the influence of two coupling terms on the 
von Neumann entropy measure using a simplified model, in which  a single excitation is considered
to propagate to neighboring sites. In reality, the electronic excitation is not confined
to a single site but distributed over all $N$ molecular sites of the crystal. The interaction
between these sites then gives rise to $N$ states, with each $N$ state characterized by a 
wavevector ${ k}$ and associated energy level. The superposition of several
such states give rise to an excitation which  propagates from one site another. This highlights
the inherent  difficulties associated with a full  computation of  von Neumann entropy of 
a realistic system. One has to incorporate the possibility that each site
becomes a  source of excitation that propagates, while  receiving 
an amplitude (of the form  in  Eq.(\ref{econ1}))
from other sites. A full  superposition of amplitudes from all possible sites have to 
be then taken into account with time as an added dimension to obtain an accurate picture
of the excitation dynamics. In order to simplify analysis
of the full model where all sites have equal probability of being excited, we 
consider the   Wootters  concurrence \cite{Woo} which
provides a convenient means to study bipartite entanglement
in the next Section. 

\section{\label{cycle} Molecular cooperativity as  a paradigm of Collective entanglement}

A notable optical response exhibited by one-dimensional 
excitonic systems is their enhanced rate of spontaneous
emission compared to that of the monomer, an 
effect known as superradiance \cite{jap1,jap2,knos}.
In a series of works \cite{shaun0,shaun1,shaun2,shaun3}, Mukamel and coworkers have 
shown a connection between 
cooperative emission mechanisms such as superradiance 
 and ``molecular cooperativity", a term used
to describe the effective coherent dynamics of a system of delocalized Frenkel excitons.
 The concept of a group of molecules acting in tandem to produce
enhanced features in $\chi^{(3)}$, the third order optical response of one dimensional
molecular systems was also shown in an earlier work by Hanamura and coworkers \cite{jap2}. 
The oscillator strength of the $K=0$ exciton state is directly proportional
to the number of molecular sites in the absence of decoherence mechanisms,
which implicates a scenario where all molecular dipoles act collectively in phase
to produce giant oscillator strengths. These results
highlight the potential in harnessing
$\chi^{(3)}$ as a quantitative measure of  multipartite entanglement 
 if a quantum-information perspective is attached to the observed
 ``cooperative" phenomena. This connection will be further examined in 
 Sec. \ref{opt}. In this section however, we  utilize a popular 
 entanglement measure, Wootters  concurrence \cite{Woo} to interpret  
superradiance of one-dimensional excitonic systems. 

The well-known  Wootters
 concurrence \cite{Woo} $C$ provides a measure of entanglement between
a pair of qubits, it is zero for separable states
and is equal to one for maximally entangled states such as
the Bell states.  $C$  has  a simple form
The average pairwise concurrence $\< C\>$ \cite{laks,imre}
associated with a system of single-particle states is given by 
\be
\label{avcon}
\< C\> = \frac{2}{N(N-1)} \left[ {\zeta}-1 \right ]
\ee
$\zeta$ is a measure of delocalization and is 
also known as inverse participation ratio (IPR) \cite{but}. 
Small values of $\zeta$ correspond to localized states
while larger values are linked with delocalization.
At the extreme limit of localization at a single site $\zeta=1,\; \< C \>=0$,
while for  states which are delocalized completely, $\zeta=N$
which yields the maximum entanglement of $\frac{2}{N}$.
For  extended excitonic system of J-aggregate 
structures, we consider  $\zeta$ as playing similar role as $N_c$ ,
the characteristic coherence size. $N_c$ has been  formulated in 
terms of the density matrix as \cite{shaun3}
\be
\label{stime}
N_c =\left[ N \sum_{m n} |\rho_{m n}|^2 \right ]^{-1}
\left(\sum_{m n} |\rho_{m n}| \right)^2 
\ee
where $N$ is the total number of molecules or monomers
and $\rho_{m n}$ is the reduced density matrix associated with sites
$m$ and $n$.  
$N_c$ provides an approximate measure by which  exciton at site $n$ is entangled with  site $m$,
and decays monotonically as a function of $(n-m)$.
In the absence of any entanglement with extreme 
localization, $\rho_{m n} = N^{-1} \delta_{nm}, \; N_c=1$ and the scaled
 concurrence obtained by dividing Eq.(\ref{avcon}) with $\frac{2}{N}$,
$\< C \>_s=0$. In the limit of complete delocalization, 
$\rho_{m n} = N^{-1}, \; N_c=N$ which yields the maximum
$\< C \>_s=1$. Thus  the inverse participation ratio 
$\zeta$ corresponds well with the characteristic coherence size $N_c$.

Using a coarse grained
approximation (CGA) (described in Appendix B of Ref.~\cite{shaun2})
to solve the exciton Green's function (similar to  Eq.(\ref{g2})), Spano et al \cite{shaun2}
determined the coherence size $N_c$ using  a microscopic 
definition  $G(0,t) \sim \exp(-N \; \gamma_r t)$.  
The radiative rate is given by $N \; \gamma_r$ in the superradiant limit at which 
excitation associated with individual sites act collectively
on the molecular aggregate. In particular, Spano et al \cite{shaun2}
obtained an empirical relation relating $N_c$ and the coupling function
$F({ k, q})$, temperature $T$, nearest neighbor transfer energy $V$ and
phonon energy,   $ \hbar \omega({ q})$. By rewriting  the  empirical relation obtained 
by Spano et al \cite{shaun2} in terms 
of the dimensionless constants $b$ and $c$ (given below
 Eq.(\ref{pron2})), we obtain
\be
\label{spano}
N_c  \approx 2.16 
\left( \frac{V^2 \; \hbar \omega({ q})}{F^2({ k, q}) \; k_B T} \right )^{\frac{1}{3}}
=  2.16 
\left( \frac{c^2}{b^2 t} \right )^{\frac{1}{3}}
\ee
where the  dimensionless temperature term, 
$t_k = \frac{k_B T }{\hbar  \omega({ q})}$.  
The molecular cooperativity due to 
the monomer's dipoles lining up collectively in phase, results in a large 
net amplitude,  giving rise to superradiant
decay \cite{shaun0} and giant nonlinear optical  properties \cite{knos}. 
The  coherence size of the associated J-band  is generally taken as the
ratio of its radiative decay rate to that of a single monomer, however  
off-diagonal elements  of the exciton density matrix in   Eq.(\ref{exdensity})
can be also used as basis to compute the coherent size.
Results obtained by Spano et al \cite{shaun2} showed pronounced decrease in 
$N_c$ at resonant points given by 
\be
\label{res}
N_c  \approx 4 \pi c \quad \quad \quad \quad \quad \quad {\rm resonance \quad points}
\ee
 when acoustic phonon frequencies
match those between the $K$th exciton level and $K=0$ exciton level.
The significance of these resonant points is attached to 
 a transfer of  oscillator strength away from  the
 $K=0$  state  due to exciton-phonon scattering. 

\begin{figure}[h!]
\hspace*{-7.5mm}
  \begin{center}
\includegraphics[width=7.0cm]{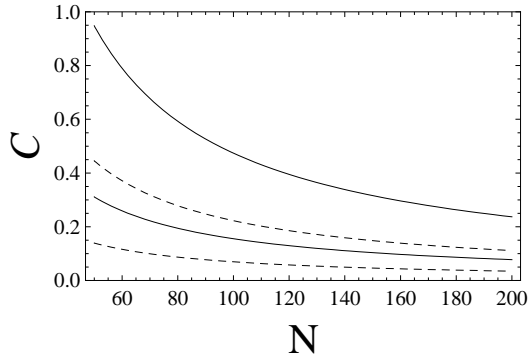}
  \end{center}
\caption{\label{w5}
Concurrence $C$ versus aggregate size $N$ at $t_k=2$.
Solid  curves for  $c=$ 15,  correspond to (from top to bottom) $b=0.5,0.1$.
Dashed  curves for  $c=$ 5,  correspond to (from top to bottom) $b=0.5,0.1$.
}
\end{figure}

The results obtained using  Eq.(\ref{spano})
in terms of the concurrence measure given in  Eq.(\ref{avcon}) 
is  illustrated in Fig.~\ref{w5}. The figure 
shows  gradual decrease in concurrence
 $C$ with aggregate size $N$ for various values of $b, c$ at constant $t_k=2$.
 The temperature dependence of $N_c$ in  Eq.(\ref{spano})
accounts for the sensitivity of  lifetimes of the J-aggregate at increasing
temperatures.  Increasing the coupling term $F({k, q})$ (or $b$)
 leads to destruction of intermolecular cooperativity, hence there is 
associated decrease in concurrence $C$ as shown in Fig.~\ref{w5}. The  superradiance phenomenon 
is thus less likely to occur at large $b$ values.
We note that   higher concurrence $C$ is obtained  in systems
with faster propagation of excitation (higher $c$ values) as also shown in Fig.~\ref{w5}.
It is imporant to note that we have restricted consideration to only
the pair-wise concurrence $C$ here, and not 
multipartite entanglement, which will be considered in the next section.

\section{\label{opt}  Theory of Two-exciton states}

In this section, we examine the  optical features of 
J-aggregates which present an experimentally   
quantifiable measure of  multipartite entanglement, in the presence of  
dephasing mechanisms \cite{briggy}. In this context, the class of 
entanglement quantity known as the Wei-Goldbart 
geometric measure of entanglement \cite{Wei2} may be convenient 
in analysing  multipartite entanglement 
of J-aggregates based on  properties of the  third order optical susceptibility, 
$\chi^{(3)}$. An important feature related to non-linear optical 
properties of material systems is the existence of one- and  two-exciton states.
One-exciton states are easily identified as states  in which 
an  excitation is present at any one of the $N$ monomers while in  a
two-exciton state, any two of  $N$ monomers are excited while the other monomers
remain in the ground state. We emphasize that the two-exciton state here is 
vastly different from the biexciton state which consists of two coupled exciton 
pair system.

Two-exciton states may be identified with entangled states
formed due to superposition of two spatially uncoupled excitons, such
states having  twice the bandwidth of the usually considered one-exciton states.
This can be shown by writing the diagonal Hamiltonian 
of the two-exciton state in $ K$ space (without involvement of phonon operators) as
\bea
\label{exHk2}
\hat{H}_{ex} &=& \sum_{k}  E_1(k) B_{1,k}^\dagger B_{1,k}+ E_2(k) B_{2,k}^\dagger B_{2,k}+ T_{k}(i j)
\\ \nonumber
T_{k}(i j)&=&\sum_{ m \neq n} \exp \left[i{ k.(m i-nj)} \right] V_{m n}
\eea
where $E_1(k)$ is energy of the exciton (labelled by subscript $1$) with  wavevector $k$
and $E_2(k)$ is energy of the second exciton (labelled by subscript $2$) with  wavevector $k$.
The exciton operators $B_{1,k}^\dagger, B_{1,k}$ retain the same meaning as used in earlier sections.
$V_{m n}$ denotes the  interaction energy for coupling of transition moments  for molecules
at $m,n$. Eq.(\ref{exHk2}) can be diagonalized further in terms of the exciton operators 
\bea
\label{exopd}
\chi_a(k) &=& \cos \beta(k)  B_{1,k}^\dagger + \sin \beta(k)  B_{2,k}^\dagger
\\ \nonumber
\chi_b(k) &=& -\sin \beta(k)  B_{1,k} + \cos \beta(k)  B_{2,k}
\eea
we obtain 
\be
\label{exH3}
\hat{H}_{ex} = \sum_{k} E^*_a(k) \chi_a(k)^\dagger \chi_a(k)+ E^*_b(k) \chi_b(k)^\dagger \chi_b(k)
\ee
where $ E^*_a(0)$ and  $E^*_b(0)$ are exciton energies associated with Davydov branches at $k=0$.
The asterisk in the superscript denotes energies which are split due to the intermolecular interaction.
Eq.(\ref{exH3}) shows that two-exciton states are highly entangled with  the degree of entanglement
determined by $\cos \beta(k)$  and $\sin \beta(k)$. 
We next make a connection between two-exciton states and generalized 
symmetric states via the third order optical susceptibility, 
$\chi^{(3)}$.

\subsection{\label{op}  Third order optical susceptibility, 
$\chi^{(3)}$ and entropy measure}

The third order optical susceptibility, 
$\chi^{(3)}$ can be written as \cite{knos}
\be
\fl
\label{sus}
\chi^{(3)}(-\omega_s;\omega_1,\omega_2,\omega_3) =
\sum \left[ \chi^{(3)}_c(-\omega_s;\omega_1,\omega_2,\omega_3) +
\chi^{(3 \star)}_c(-\omega_s;\omega_1,\omega_2,\omega_3)
\right ]
\ee
where the summation is performed over all possible permutations
of frequencies of incident light, $\omega_1,\; \omega_2, \; \omega_3$ 
and $\omega_s = \sum_{i} \omega_i$. Expressions for  
$\chi^{(3)}_c(-\omega_s;\omega_1,\omega_2,\omega_3)$ and its conjugate 
are lengthy and  given in an earlier work
 by  Knoester \cite{knos}, 
in terms of $\mu_{0;e1}$ and $\mu_{e2;e2,e3}$.  $\mu_{0;e1}$ is the  dipole moment
associated with transition between the ground state and one-exciton state in which
an excitation is present at any one of the $N$ monomers, 
a generalization of 
the  well known Dicke state 
$|W \> =\frac{1}{\sqrt{N}} \left(|100...0\>+|01...0\>+...+|0...01\> \right)$. 
$\mu_{e2;e2,e3}$ is the  dipole moment
associated with transition between the one-exciton state  and 
the two-exciton state. Using Eq.(\ref{exH3}) as basis, 
we consider that the two-exciton state are a manifestation of  
symmetric states,  which are dependent on the 
number of ground states or $0$'s \cite{Wei1,Wei2}
\be
\label{weis}
|S(N,M)\rangle\equiv \sqrt{\frac{M!(N-M)!}{N!}}
\sum_{\rm{\scriptstyle permutations}}
|0\>^{\otimes M}\otimes|1\>^{\otimes {(N-M)}}
\ee
The states in  Eq.(\ref{exH3}) are known to have 
an entanglement entropy measure \cite{Wei1,Wei2} 
\bea
\label{wei1}
 E_{\Lambda}(N,M) &=& - \ln \Lambda(N,M)^2 \\ 
\label{wei2}
\Lambda(N,M) &=& \sqrt{\frac{N!}{M!(N\!-\!M)!}}
\left(\frac{M}{N}\right)^{\frac{M}{2}}
{\left(\frac{N-M}{N}\right)}^{\frac{N\!-\!M}{2}}.
\eea 
For fixed $N$, the minimum entanglement
eigenvalue or geometric measure of entanglement \cite{Wei2}
 $\Lambda_{\max}$ occurs for 
$M=N/2$ (even $N$) and $M=(N\pm1)/2$ (odd $N$). 

 We consider that 
transitions involved in nonlinear optical effects occur via a 
simplified pathway represented as follows
\bea
\label{nonlin}
|S(N,0)\rangle\; && \longrightarrow \; |S(N,1)\rangle \; ({\rm one-exciton \; state})
\\
\nonumber && 
 \; \longrightarrow \;|S(N,2)\rangle \; ({\rm two-exciton \;  state})
\eea
The  terms of the form given in Eq.(\ref{weis}) 
are expected to give rise to the observed
dependence of $\chi^{(3)}$ on non-local effects associated with 
energy transfer within various exciton site basis.
Accordingly $\chi^{(3)}$ scales with the entanglement entropy $E_{\Lambda}$ in Eq.(\ref{wei1}),
however the explicit nature of relation between the two quantities is 
 not immediately clear, this predicament also applies to
a possible  relation  between $\chi^{(3)}$ and  genuine multipartite
entanglement involving all $N$ monomers. Here we keep the analysis tractable by
examining the dependence of the intensity of
 third-harmonic generation (THG) $I_{THG} \propto 
|\chi^{(3)}(-3 \omega;\omega,\omega,\omega)|^2$ on $N$ instead. We simplify
the approach via the following proposed relations  
\bea
\label{sim}
\mu^2_{0;e1} &\approx& N \mu^2  \zeta_1\\
 \nonumber
\mu_{e2;e2,e3}^2 &\approx&   N \mu^2 \zeta_2 \\  \nonumber
\zeta_1 &=&  \frac{E_{\Lambda}(N,1)}{E_{\Lambda}(N,N/2)} \\  \nonumber
\zeta_2 &=&  \frac{E_{\Lambda}(N,2)}{E_{\Lambda}(N,N/2)} 
\eea
where $\mu$ is the transition dipole of individual monomers and
the entanglement entropy measure $E_{\Lambda}(n,k)$ is given in 
Eq.(\ref{wei1}). 
$\zeta_1$ and $\zeta_2$ are entropy measures
associated with the transition efficiencies of one-exciton state 
and two-exciton state respectively.

\begin{figure}[h!]
  \begin{center}
\includegraphics[width=7.0cm]{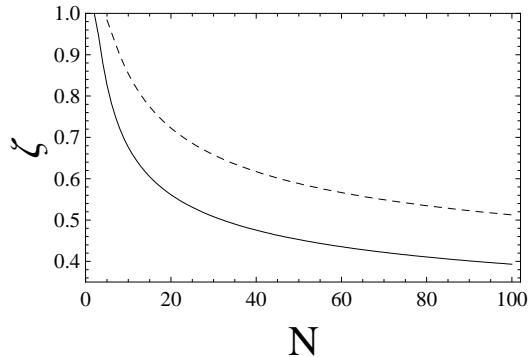}
  \end{center}
\caption{\label{w6}
Entropy measures  $\zeta_1$ (solid line) and $\zeta_2$ (dotted) versus 
aggregate size $N$. 
}
\end{figure}

\begin{figure}[h!]
\begin{center}
\includegraphics[width=8.0cm]{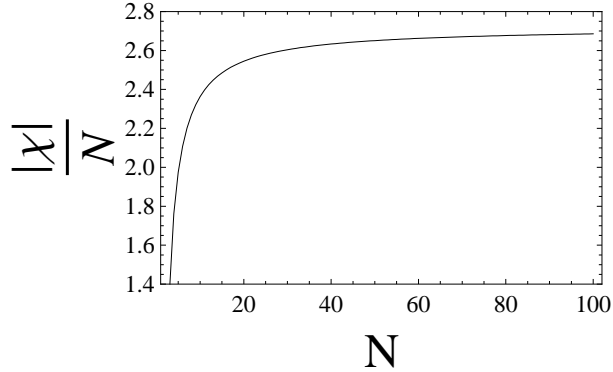}
\end{center}
\caption{\label{w7}
$\frac{1}{N}|\chi^{(3)}(-3 \omega;\omega,\omega,\omega)|$ ( see Eq.(\ref{suscep}))
 versus aggregate size $N$. 
}
\end{figure}

Fig.~\ref{w6}  
shows  gradual decrease in these entropy measures with increasing
aggregate size $N$, with $\zeta_2$  the entropy measure associated
with two-exciton band seen as more resilient, which is expected
as the associated entanglement persists for longer times due its
wider bandwidth. In both cases, the entropy decreases with $N$ due
to the faster increase of  $E_{\Lambda}(N,N/2)$ with $N$.
Using Eq.(\ref{sim}) and  $\omega = \frac{1}{3}\Delta E$ where the excitation 
energy $\Delta E$ is defined in Eq.(\ref{exH}),
we write the absolute value of the third order optical susceptibility with 
damping rate $\Gamma$ as 
\be
\label{suscep}
|\chi^{(3)}(-3 \omega;\omega,\omega,\omega)|= N  
\frac{\mu^2 E_{\Lambda}(N,1) E_{\Lambda}(N,2)} {2 \Gamma \hbar^3 \;(\omega^2- \Delta E^2)}
\ee
Fig.~\ref{w7}  
shows  gradual increase of
 $\frac{1}{N}|\chi^{(3)}(-3 \omega;\omega,\omega,\omega)|$ (in which
a constant term in Eq.(\ref{suscep}) was factored out) 
with aggregate size $N$ before reaching an almost constant value which is independent
of the size of system. The results of Fig.~\ref{w7}  
illustrate the important fact that properties of the J-band 
is determined by a {\it critical number  of coherently
coupled} and entangled monomers rather than the total number $N$
of monomers present in the system. 

The slope in Fig.~\ref{w7}  
yields  the degree
of entanglement or multipartite entanglement measure.
In this respect, the {\it onset of zero slope} during  change
of third order optical susceptibility with the system size $N$ 
can be  viewed as the point of maximum  entangled state.  
The results in Fig.~\ref{w7} 
are consistent with earlier results \cite{knos} which showed the critical
role played by the exciton delocalization length even Knoester \cite{knos} 
did not analyse his results from a quantum entanglement perspective.

The results obtained in  Fig.~\ref{w7}  show that  scaling  of  the third order optical response
$\chi^{(3)}$ with exciton delocalization size may
be used as basis in quantifying multipartite entanglement 
in J-aggregates or systems composed of excitonic spin chains. The result obtained here 
is significant considering 
that a rigorous definition of genuine multipartite
entanglement is still lacking in the literature. Thus 
entanglement measures may be  defined based on
experimental observations and taking into account decoherence 
mechanisms \cite{tiop1,tiop2},  exciton-photon interactions and laser detuning processes.
In this regard, the results obtained so far has established some basic
guidelines for an experimentally   demonstrable  measure of identifying multipartite entanglement
in excitonic systems.
In the table below, we list various quantum information processing attributes in 
 J-aggregates systems.
\bigskip
\begin{center}
{{\bf Table 1} \; {\it Quantum information processing attributes in 
 J-aggregates}}
\bigskip

\begin{tabular}{|c|c|c|} \hline
{\bf  State/Event} & {\bf Quantum information attributes} \\ \hline
One-exciton state &  Encompasses Dicke State \\ \hline
 Two-exciton states & Encompasses Symmetric States\\ \hline
Molecular cooperativity & Paradigm of collective entanglement \\ \hline
J-aggregate systems & Channels of quantum information/energy propagation \\ \hline
Third order optical susceptibility & Viable multipartite entanglement measure\\ \hline
\end{tabular}
\end{center}
\bigskip

In conclusion, we have  demonstrated the usefulness of the 
von Neumann entropy,  Wootters
 concurrence and geometric measure of entanglement
in characterizing the  quantum information processing attributes
quantum J-aggregates. We identify two-exciton states (entangled states
 of  uncoupled excitons)  and molecular cooperativity 
(manifestation of multipartite entanglement) as features which can be exploited
for information processing. The  entanglement and quantum correlations 
associated with a delocalized exciton over many monomer sites  provides
a clearer understanding of the processes behind 
observed behavior of  J-aggregates. The delocalization 
of electronic states which give rise to  narrowness
of  J-aggregate spectrum   as expected is complicated by
the fine interplay between  excitation transfer mechanism and 
decoherence associated with a background of phonon bath. With control 
of the  dispersive coupling ($\chi({ q})$) and  
 resonance coupling ($F({ k, q})$) parameters  via temperature and matrix characteristics,
 J-aggregate systems may be utilized as robust channels for 
large scale energy propagation.

We have also demonstrated that  scaling 
 of  the third order optical response
$\chi^{(3)}$ with exciton delocalization size 
provides an experimentally   demonstrable  
measure of quantifying  multipartite entanglement
in  Frenkel excitonic systems. $\chi^{(3)}$
may be used to distinguish 
between system with low and high fidelities. 
In future works, we seek to address
the question of whether the J-band is a  maximally entangled state 
or one that exhibits genuine multipartite. 
Lastly, the results of this work will be useful in  understanding 
 many-body correlated dynamics of excitonic systems, 
with  impact on future experimental work involving measurement
of multipartite entanglement. This is of considerable significance 
as a rigorous definition of genuine multipartite
entanglement still remains an open question.

\section*{\label{ref}References}


\begin{thebibliography}{9}


\bibitem{jelly}
E.E. Jelley, Nature {\bf 138}, 1009  (1936).

\bibitem{sch}
G. Scheibe, Angew. Chem. {\bf 50}, 212 (1937). 

\bibitem{brigg1}
J. S. Briggs and  A. Herzenberg, Mol. Phys. {\bf 21}, 865 (1971).


\bibitem{brigg2}
J. Roden, A. Eisfeld, W. Wolff, and W. T. Strunz,
 Phys. Rev. Lett. {\bf 103}, 058301 (2009).


\bibitem{nel}
M. A. Nielsen and I.L.  Chuang,
{\it Quantum Computation and Quantum Information}
(Cambridge University Press, Cambridge, U.K., 2000).

\bibitem{Dav}
A.S. Davydov, {\it Theory of Molecular Excitons} (Plenum, New York, 1971).


\bibitem{Craig}
D. P. Craig and S. H. Walmsley, {\it Excitons in Molecular Crystals} (Benjamin Inc.,
New York, 1968).

\bibitem{toy}
Y. Toyozawa, {\it Optical Processes in Solids} (Cambridge, New York, 2003).


\bibitem{expt1}
F. Milota, J. Sperling, A. Nemeth, and  H. F. Kauffmann,
Chem. Phys. {\bf 357}, 45  (2009).


\bibitem{expt2}
I. Stiopkin, T. Brixner, M. Yang, and G.R. Fleming, 
J. Phys. Chem. B {\bf 110}, 20032 (2006).

\bibitem{micr}
P. Schouwink, H. V. Berlepsch, L. D\"ahne, and R. F. Mahrt,
Chem. Phys. Lett. {\bf 344}, 352 (2001).

\bibitem{sp}
W. T. Simpson and D. L. Peterson, J. Chem. Phys. {\bf 26}, 588 (1957).


\bibitem{Klaf}
J. Klafter and J. Jortner, Chem. Phys. Lett. {\bf 50}, 202 (1977).


\bibitem{Knapp}
E. W. Knapp, Chem. Phys. {\bf  85}, 73 (1984).



\bibitem{spookE}
A. Einstein, B Podolsky, N. Rosen, Phys. Rev. {\bf 47}, 777 
(1935).

\bibitem{spookJ}
K Yokota, T Yamamoto, M Koashi and N Imoto, 
New J. Phys. {\bf 11}, 033011  (2009). 

\bibitem{tiop1}  A. Thilagam, 
J. Phys.: Condens. Matter {\bf  21}, 045504 (2009).

\bibitem{tiop2} A. Thilagam and M. A. Lohe, 
J. Phys.: Condens. Matter {\bf  20}, 315205 (2008).

\bibitem{tiop3} A. Thilagam ,
J. Phys. A: Math. Theor. 43 155301 (2010) 

\bibitem{briggy}
A. Eisfeld and J. S. Briggs, Phys. Rev. Lett. {\bf 96}, 113003
(2006).


\bibitem{ThilaPRA}
A. Thilagam, Physical Review A {\bf 81}, 032309 (2010).


\bibitem{scott}
A. J. Scott, Phys. Rev. A {\bf 69}, 052330 (2004).

\bibitem{hor}
R. Horodecki, P. Horodecki, M. Horodecki, and K. Horodecki,
Rev. Mod. Phys., {\bf  81}, 865 (2009).

\bibitem{ami}
L. Amico, R. Fazio, A. Osterloh, and V. Vedral, Rev. Mod.
Phys. {\bf  80}, 517 (2008).


\bibitem{hein}
 M. Hein, W. D\"ur, J. Eisert, R. Raussendorf, M. Van den Nest, and H. J. Briegel.
{\it Entanglement in graph states and its applications}, Proc. of the Int. School of
Physics "Enrico Fermi" on "Quantum Computers, Algorithms and Chaos", July
2005. quant-ph/0602096

\bibitem{plen}
M. B. Plenio and S. Virmani, Quantum Inf. Comput. {\bf  7}, 1
(2007).

\bibitem{eise}
J. Eisert and D. Gross, {\it Lectures on Quantum Information},
eds. D. Bruß and G. Leuchs (Wiley-VCH, Weinheim,
2007).


\bibitem{coff}
V. Coffman, J. Kundu and W. K. Wootters,
Phys. Rev. A {\bf61}, 052306 (2000).

\bibitem{kit}
A. Kitaev and J. Preskill, Phys. Rev. Lett. {\bf 96},
110404 (2006).


\bibitem{Woo} W. K. Wootters, Phys. Rev. Lett. {\bf 80},  2245 (1998).


\bibitem{Wei2}
T. C. Wei and P. M. Goldbart, Phys. Rev. A {\bf 68}, 042307(2003).



\bibitem{suna}
A. Suna, Phys. Rev. {\bf 135}, A111 (1964).


\bibitem{craigdiss}
D. P. Craig and L. A. Dissado, Chem. Phys.  {\bf 14}, 89 (1976). 


\bibitem{econ}
E. N. Economou, {\it Green's Functions in Quantum Physics}
(Springer-Verlag, Berlin, 1979).

\bibitem{merry}
R. E. Merrifield, J. Chem. Phys. {\bf 28}, 647 (1958).


\bibitem{paw}
G. S. Pawley and S. J. Cyvin, J. Chem. Phys. {\bf 52}, 4072 (1970)



\bibitem{thilaIOP}
A. Thilagam, J. Phys. A: Math. Theor. {\bf  42}, 335301 (2009).


\bibitem{jap1} T. Takagahara, Surface Science, {\bf 196}, 590 (1987).


\bibitem{jap2}
T. Tokihiro, Y. Manabe, and E. Hanamura,
Phys. Rev. B {\bf 47} 2019 (1993).


\bibitem{knos}
J. Knoester, Chem. Phys. Lett. {\bf 203}, 371 (1993).


\bibitem{shaun0}
F. C. Spano, J. R. Kuklinski, and S. Mukamel,
 Phys. Rev. Lett. {\bf 65}, 211 (1990).


\bibitem{shaun1}
T. Meier, Y. Zhao, V. Chernyak, and S. Mukamel,
J. Chem. Phys. {\bf 107}, 3876 (1997).

\bibitem{shaun2}
F. C. Spano, J. R. Kuklinski, and S. Mukamel,
J. Chem. Phys. {\bf 94}, 7534 (1991).


\bibitem{shaun3}
O. Dubovsky  and S. Mukamel,
J. Chem. Phys. {\bf 95}, 7828 (1991).


\bibitem{laks}
A. Lakshminarayan and V. Subrahmanyam, Phys. Rev. A {\bf 67}, 052304 (2003).


\bibitem{imre}
I. Varga and J. A. Mendez-Bermudez,
Physica Status Solidi C. {\bf 5}, 867 (2008).


\bibitem{but}
A. Barelli, J. Bellissard, P. Jacquod, and D. L. Shepelyansky,
Phys. Rev. Lett. {\bf 77} 4752 (1996).


\bibitem{Wei1}
T. C. Wei, K. Nemoto, P. M. Goldbart, P. G. Kwiat, 
W. J. Munro, and F. Verstraete, 
Phys. Rev. A {\bf 67}, 022110 (2003).


\end{thebibliography}
\end{document}